\documentclass[apj,numberedappendix]{emulateapj}
\usepackage{ifthen}
\usepackage{amsmath}
\usepackage[alpha,z,expert]{tmsfnts}
\usepackage{graphicx}



\shorttitle{Calibration of mass--temperature relation}
\shortauthors{Pedersen \& Dahle}
\slugcomment{Revised version submitted to ApJ}

\begin{document}

\title{Calibration of the Mass--Temperature Relation for Clusters of 
Galaxies Using Weak Gravitational Lensing\altaffilmark{1}}

\author{Kristian Pedersen}
\affil{Dark Cosmology Centre, Niels Bohr Institute, University of Copenhagen, 
Juliane Maries Vej 30, DK-2100, Copenhagen \O, Denmark}
\email{kp@dark-cosmology.dk}
\author{H{\aa}kon Dahle\altaffilmark{2}}
\affil{Institute of Theoretical Astrophysics, University of Oslo, P.O. Box 1029, 
Blindern, N-0315 Oslo, Norway}
\email{hdahle@astro.uio.no}

\altaffiltext{1}{Based on
observations made with the Nordic Optical Telescope,
operated on the island of La Palma jointly by Denmark, Finland,
Iceland, Norway, and Sweden, in the Spanish Observatorio del Roque de
los Muchachos of the Instituto de Astrof\'\i sica de Canarias}

\altaffiltext{2}{Visiting observer, University of Hawaii 2.24m Telescope at 
Mauna Kea Observatory, Institute for Astronomy, University of Hawaii}

\begin{abstract}
{The main uncertainty in current determinations of the 
power spectrum normalization, $\sigma_8$, from abundances of X-ray luminous galaxy clusters 
arises from the calibration of the mass--temperature relation. We use our weak 
lensing mass determinations of 30 clusters from the hitherto largest sample of clusters with
lensing masses, combined with X-ray temperature data from the literature, to calibrate 
the normalization of this relation at a temperature of 8~keV,
$M_{\rm 500c,8 keV}= (8.7 \pm 1.6)\,h^{-1}\, 10^{14}M_{\sun}$.
This normalization is consistent with previous lensing-based 
results based on smaller cluster samples, and with some predictions from numerical
simulations, but higher than most normalizations based on X-ray derived
cluster masses.
Assuming the theoretically expected slope $\alpha=3/2$ of the mass--temperature
relation, we derive $\sigma_8 = 0.88 \pm 0.09$ for a spatially-flat $\Lambda$CDM 
universe with $\Omega_m = 0.3$. 
The main systematic errors on the lensing masses result from extrapolating the cluster masses 
beyond the field-of-view used for the gravitational lensing measurements, and 
from the separation of cluster/background galaxies,  contributing each with a scatter of
20\%{}. Taking this into account, there is still significant intrinsic scatter in the 
mass--temperature relation indicating that this relation may not be very tight,
at least at the high mass end. Furthermore, we find that dynamically relaxed clusters are 
$75\pm 40\%$ hotter than non-relaxed clusters.}
\end{abstract}


\section{Introduction} 

The abundance of massive clusters of galaxies provides sensitive constraints 
on the cosmological parameters that govern structure growth in the universe. 
However, a prerequisite for this is reliable mass measurements for large samples 
of clusters with well-understood selection criteria.  
Cluster mass measurements used have traditionally come from 
virial analysis of the velocity dispersion 
measurements of cluster galaxies (e.g., Frenk et al.\ 1990; Carlberg et al.\ 1997; 
Borgani et al.\ 1999), or X-ray temperature measurements of the hot intra-cluster gas 
under the assumption that the gas is in hydrostatic equilibrium 
\citep[for a review see ][]{rosati}. 
Satellite observatories such as ROSAT, ASCA, XMM-Newton and {\it Chandra} 
have made increasingly accurate X-ray temperature measurements of clusters, 
and have produced well-defined cluster samples of sufficient size to 
accurately measure the X-ray temperature and luminosity functions
\citep[e.g.][]{henry04,boehringer}. However, the relation between cluster mass 
and X-ray temperature and luminosity, respectively, must be determined to convert 
these into a reliable cluster mass function. 

X-ray luminosities are available for large samples of clusters, but the
X-ray luminosity is highly sensitive to the complex physics of cluster cores, 
making it challenging to relate to cluster mass.  
Measuring X-ray temperatures is observationally much more demanding, but the
X-ray temperature is mainly determined by gravitational processes and is hence
more directly related to cluster mass than X-ray luminosity. 
Both from simulations and observations the intrinsic scatter in mass around 
the mass--temperature relation is thus found to be much smaller 
\citep[$\Delta M/M \approx 0.15$, ][]{evrard96,borgani,sanderson,vik} than the 
scatter in mass around the mass--luminosity relation 
\citep[$\Delta M/M \approx 0.4$, ][]{reiprich}.

Two different routes have been followed for determining the mass--temperature
relation. Most studies have used a small sample (up to about a dozen) of
supposedly well understood clusters for which the assumptions underlying the
mass determination should be fulfilled to a high degree. The main concern
for this approach is that the selected clusters may not be representative of
the whole cluster population, and therefore the derived mass--temperature relation
may only apply to a subset of clusters. Alternatively, the mass--temperature
relation may be determined from a large sample of more objectively selected 
clusters. This is more fruitful when comparing such a locally determined 
mass--temperature relation to a sample of high-redshift clusters where the data 
quality does not allow a similar selection of the ``most suitable'' clusters. 
Also, mass--temperature relations derived from simulations are usually
based on a large range of simulated clusters with no pre-selection. 
Hence it is most appropriate to compare observationally obtained 
mass--temperature relations determined from all available clusters to 
the relations from simulations.
On the other hand, for some of the clusters in such a sample the hydrostatic
assumption may be invalid, making X-ray based mass determinations unreliable  
for a subset of the clusters. A larger scatter (which may not be symmetric) around
the mean mass--temperature relation may be expected, when such clusters are 
included.

There are still poorly understood systematic uncertainties 
associated with establishing the mass--temperature relation.
The normalization of the mass--temperature relation based on cluster masses
determined from X-ray data \citep[][]{fin,vik,arnaud05}, tend to differ significantly 
between studies, and from the expectations based on numerical 
simulations (e.g., Evrard et al.\ 1996; Eke et al.\ 1998; Pen 1998; 
Borgani et al.\ 2004). 
The determination of this normalization is currently the dominating
source of discrepancies between the reported values for the power spectrum 
normalization on the scale of galaxy clusters, 
$\sigma_8$, derived from the observed cluster temperature function
(Huterer \& White 2002; Seljak 2002; Pierpaoli et al.\ 2003; Henry 2004). 
Observations using X-ray based mass determinations have traditionally favored 
low normalizations (and hence low values of $\sigma_8$), 
while simulations have favored somewhat higher normalizations.       

Gravitational lensing provides an opportunity to measure cluster masses without 
invoking the assumption of hydrostatic equilibrium in the hot intra-cluster gas 
implicit in the X-ray based mass determinations. Also, in this case the measurement 
of cluster mass is truly independent of the X-ray temperature measurement.
Hjorth, Oukbir \& van Kampen (1998) used weak gravitational mass measurements 
for eight clusters drawn from the literature to find a relation between mass over 
cluster-centric radius, and temperature. They determined a normalization of this relation
consistent with the value predicted by Evrard et al.\ (1996), but with a preference for
somewhat higher cluster masses (if the redshift scaling of equation~\ref{eq:theoryMT} is 
assumed, see below).  
However, Smith et al.\ (2005; hereafter S05) determined a mass--temperature relation with a 
normalization significantly lower than indicated by the 
Hjorth et al.\ (1998) study. 
S05 based their results on a sample 
of 10 clusters with weak lensing masses and temperatures determined from {\it Chandra} data.

Here, we present a new weak gravitational lensing-based measurement of the 
normalization of the mass--temperature relation. The main improvements with 
respect to the work of Hjorth et al.\ (1998) and S05 is that we
use a significantly larger cluster sample which represents a 
significant fraction of all the clusters in an even larger sample with  
well-defined objective selection criteria (Dahle et al.\ 2002; 
H.\ Dahle 2006, in preparation). An additional improvement over the work of 
Hjorth et al.\ (1998) is that the weak lensing analysis has been 
performed in a consistent way for all clusters, using the same shear estimator and 
making the same assumptions about e.g., the typical redshift of the lensed galaxy 
population and the degree of contamination by cluster galaxies. 
We note that the early data set of clusters with 
published weak lensing masses used by Hjorth et al.\ (1998) is biased 
at some level towards systems that were observed 
because of ``extreme'' properties, such as being the hottest or most X-ray 
luminous system known at the time, or having a large number of 
strongly gravitationally lensed arcs.  
Furthermore, we note that our gravitational lensing measurements are made at larger 
radii than probed by S05, requiring smaller extrapolations to estimate the mass within 
e.g., the virial radii of the clusters. 

The data set used for the analysis is described in \S~\ref{sec:data}, 
our results for the mass--temperature relation and $\sigma_8$ are presented 
in \S~\ref{sec:results}, and our results are compared to other work and the
implications discussed in \S~\ref{sec:discussion}. 

Except when specifically noted otherwise (for easy comparison to previous results
using different cosmologies), we assume a spatially--flat cosmology 
with a cosmological constant ($\Omega_m = 0.3$, $\Omega_{\Lambda} = 0.7$), and 
the Hubble parameter is given by $H_0 = 100 h\, {\rm km}\, {\rm s}^{-1} {\rm Mpc}^{-1}$.

\ \\ \

\section{Data Set}
\label{sec:data}

\subsection{Weak lensing data} 
\label{sec:wldata}

Our weak lensing data set is a sample of 30 clusters (see Table~\ref{tab:dataset}), 
of which 28 were included 
in the weak lensing cluster sample of Dahle et al.\ (2002). Data for 
two additional clusters come from a recent extension of this data set (H. Dahle 2006, 
in preparation). 
The clusters targeted for these weak lensing studies were generally selected to 
lie above an X-ray 
luminosity limit $L_{X, 0.1 - 2.4\,{\rm keV}} \geq 6 \times 10^{44}$ ergs s$^{-1}$
(this luminosity limit is for our chosen cosmology with $h = 0.7$) and within a 
redshift range 
$0.15 < z_{\rm cl} < 0.35$. The observed clusters were selected from the X-ray luminous 
cluster samples 
of Briel \& Henry (1993) and Ebeling et al.\ (1996;1998;2000). The cluster samples 
of the first two of these papers 
are based on correlating an optically selected cluster sample 
(Abell 1958; Abell, Corwin, \& Olowin 1989) with X-ray sources from the 
ROSAT All-Sky Survey (RASS; Tr{\"u}mper et al.\ 1993), while the two last papers 
contain X-ray flux limited cluster catalogs, also based on RASS. Of the total sample 
of 30 clusters, three (\objectname{A959}, \objectname{A1722}, and \objectname{A1995}) are drawn from the Briel \& Henry (1993) 
sample and two (\objectname{A209} and \objectname{A2104}) are drawn from the XBAC sample of Ebeling et al.\ (1996). 
Of the remaining 25 clusters, 22 are included in the X-ray brightest cluster sample 
(BCS) of Ebeling et al.\ (1998), while three (\objectname{A611}, \objectname{A1576}, and \objectname{Zw3146}) come from its 
low-flux extension (eBCS; Ebeling et al.\ 2000). 
Of the BCS and eBCS clusters in our sample, 24 objects are included in a
volume-limited sample of 35 clusters selected from the BCS and eBCS samples (Dahle 2006). 
Hence, while our sample is not strictly physically well-defined (in the sense that the availability of an 
X-ray temperature measurement is one of the defining selection criteria), it still has significant 
overlap with a well-defined cluster sample. In a recent paper, Stanek et al.\ (2006) discuss how a significant scatter 
around the mean mass-luminosity relation may cause a significant Malmquist bias in X-ray flux-limited cluster samples, 
causing high-mass, low flux clusters to drop out at high redshifts. This would result in a bias in the 
mass-luminosity (or mass-temperature) relation derived based on such a sample. We note, however, that 
although the RASS-based samples from which our cluster sample is drawn are flux-limited, the cluster sample discussed 
here quite closely approximates a volume-limited sample, and we therefore expect any such bias to be 
negligible.   
     
The observations were made with the $8192^2$ UH8K mosaic CCD camera and the $2048^2$ Tek CCD 
camera at the 2.24m University of Hawaii Telescope and 
with the $2048^2$ ALFOSC CCD camera at the 2.56m Nordic Optical
Telescope. All clusters were imaged in both the $I$- and $V$-band, 
with typical total exposure times of 3.5h in each passband for the UH8K data and 
$\sim 1.5$h for the data obtained with the more sensitive $2048^2$ detectors. 
The seeing was in the range $0\farcs6 \leq {\rm FWHM} \leq 1\farcs1$ for all the imaging data 
used for the weak lensing analysis. The median seeing was 0\farcs82 in the $I$-band and 
0\farcs9 in the $V$-band. This gave typically $\sim 25$ usable background galaxies per 
square arcminute, or a ``figure of merit'' value of 
$\sum Q^2/d{\Omega} \simeq 1.5 \times 10^5$deg$^{-2}$, as defined by Kaiser (2000).
As noted below, the background galaxies were selected based on signal to noise ratio rather 
than magnitude, with limits corresponding to $21 \lesssim m_I \lesssim 24.5$ and
$22 \lesssim m_V \lesssim 25.5$ for point sources.   
The observations and data reduction of the data set used for the weak lensing 
mass measurements are described in detail by Dahle et al.\ (2002). 

Major efforts are being made to improve the methods for the estimation of weak 
gravitational lensing, particularly in connection with ongoing and future studies of 
``cosmic shear'' based on wide-field optical surveys. The requirements for the precision of shear 
estimates in these surveys are substantially more stringent than for weak lensing 
observations of massive clusters, given the significantly weaker lensing effects measured 
in random fields. 

In this work, we have used the  
shear estimator of Kaiser (2000), which was ``blind-tested'' (along with several 
other shear estimators) by Heymans et al.\ (2005), using simulated lensing data.   
The shear estimator of Kaiser (2000) is more mathematically rigorous than the currently 
most widely used shear estimator (Kaiser, Squires, \& Broadhurst 1995), 
but it displays a significant 
non-linear response to shear, unlike most other shear estimators. 
If we correct our shear values using a second order polynomial based on the test 
results of Heymans et al.\ (2005), we find that most cluster masses stay within $+/-15$\% 
of the mass 
calculated based on uncorrected shear values. Furthermore, the change in 
average cluster mass is $<2$\%, i.e., there is very little systematic shift in mass. 
In the end, we chose not to apply this correction, since it would, in a few cases, require
extrapolations outside the range of shear values over which the shear estimator has 
been tested.   
For more details about the practical implementation of this shear estimator, see  
Dahle et al.\ (2002). 

\begin{figure}
\resizebox{8cm}{!}{\includegraphics{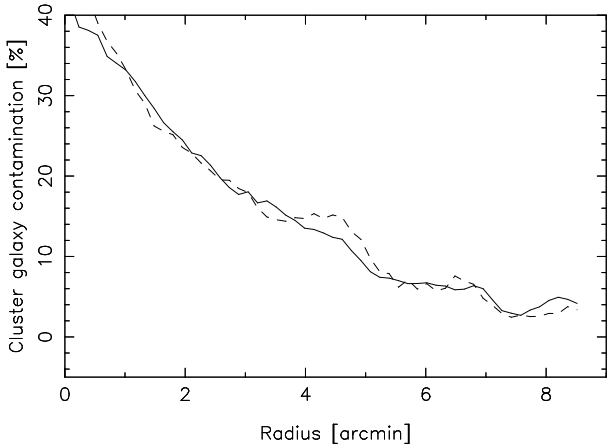}}
\caption[Cluster galaxy contamination in the faint galaxy samples.]
{Cluster galaxy contamination in the faint galaxy catalogs as a function of distance from the cluster center. The solid line represents an average of 6 clusters at an average redshift $\langle z \rangle=0.31$, while the dashed line represents an average of 5 clusters at $\langle z \rangle =0.23$.}
\label{fig:cluscont}
\end{figure}

To convert the measurements of weak gravitational shear into actual cluster masses, 
the distances to the background galaxies need to be known. 
The background galaxy redshifts were estimated from spectroscopic and photometric 
redshifts in the Hubble Deep Field (for details, see Dahle et al.\ 2002). 
For our data set and chosen cosmological model, the average value of the ratio 
between the lens-source and observer-source angular diameter distances, 
$\beta \equiv D_{ls}$/$D_s$, is well approximated by the relation 
$\langle \beta \rangle = 1.37 z_{\rm cl}^2 - 2.00 z_{\rm cl} + 1.01$ within 
the redshift range of our cluster sample. This then provides an effective 
critical surface density for lensing  
($\Sigma_{\rm crit} = (c^2/4\pi G)(D_l \langle \beta \rangle)^{-1}$; 
where $D_l$ is the angular diameter distance to the cluster), which is used for 
deriving cluster masses from the shear estimates. The quoted value of $\langle \beta \rangle$
corresponds to the value at large cluster radii; at smaller radii a correction term has to 
be employed to account for contamination by cluster galaxies, as discussed below and illustrated 
in Figure~\ref{fig:cluscont}.  
 
The observable galaxy shape distortions caused by gravitational lensing provide a 
measurement of the reduced tangential shear, 
$g_T = \gamma_T/(1-\kappa)$, where $\gamma_T$ is the 
tangential component of the shear and $\kappa$ is the convergence.  
We fit an NFW-type mass density profile,  

\begin{equation} 
\rho (r) = \frac{\delta_c\, \rho_c (z)}{(r/r_s) (1+ r/r_s)^2} 
\label{eq:NFW}
\end{equation} 

\noindent
(Navarro, Frenk, \& White 1997), 
to the observed reduced shear profile $g_T (r)$ of each cluster.  
Here, $\rho_c (z)$ is the critical density of the universe at the redshift of 
the cluster, and  

\begin{equation} 
\delta_c = \frac{200}{3} \frac{c_{200}^3}{\ln (1+c_{200}) - c_{200}/(1+c_{200})}. 
\label{eq:deltacdef}
\end{equation} 
 
\noindent
We assumed a concentration parameter $c_{200} = c_{\rm vir}/1.194 = 4.9 / (1+z)$, 
corresponding to the median halo concentration predicted by Bullock et al.\ (2001) for a 
$M_{\rm vir} \simeq 8\times 10^{14} M_{\sun}$ cluster from 
simulations of dark matter halos in a $\Lambda$CDM universe. 
Here, $c_{200} = r_{200c} / r_s$, and $c_{\rm vir} = r_{\rm vir} / r_s$, 
where $r_{200c}$ is defined as the radius within which the average mass
density is 200 times the critical density $\rho_{c} (z)$, and $r_{\rm vir}$ 
is the virial radius of the cluster.

The lensing properties of the NFW model have been 
calculated by Bartelmann (1996) and Wright \& Brainerd (2000). 
From our fit, we calculated $M_{500c}$, the mass enclosed 
by the radius $r_{500c}$. 
The mass estimates are listed in Table~\ref{tab:dataset}.  
The shear measurements used for the fit were made at clustercentric radii 
$50\arcsec < r < 180\arcsec$ for the clusters that 
were observed with $2048^2$ CCD cameras and $150\arcsec < r < 550\arcsec$ 
for the clusters that were observed with the UH8K camera.
By comparison, we find $r_{500c}$ values typically in the range 
$300\arcsec < r_{500c} < 600\arcsec$ for 
the clusters we study here. In many cases, we need to extrapolate the 
NFW profile out to $r_{500c}$ (in Table~\ref{tab:dataset} we list the 
ratio of the outermost radii of our shear measurements, $r_{\rm fit}$, to   
$r_{500c}$, and note that $r_{500c} = 0.66 r_{200c}$ for our chosen NFW model).
In this extrapolation, we assume the median NFW concentration parameter given 
above. Hence, any intrinsic scatter in $c_{\rm vir}$ will introduce an 
extra uncertainty in the cluster mass estimates. If we assume a random scatter around 
the mean value of $c_{\rm vir}$ at the level (a $1 \sigma$ $\Delta (\log c_{\rm vir}) \sim 0.18$) 
predicted by Bullock et al.\ (2001), we find a corresponding scatter in the mass estimates 
of 20\% for our data set. This additional scatter is not included in the uncertainties of the 
listed mass measurements in Table~\ref{tab:dataset}, but is considered further in
Section~3.2.

The measured gravitational lensing signal is sensitive to the two-dimensional 
surface mass distribution, including mass associated with the cluster 
outside $r_{500c}$, and random structures seen in projection 
along the line of sight (Metzler, White \& Loken 2001; Hoekstra 2001; 
Clowe, De Lucia \& King 2004; de Putter \& White 2005).   
This will introduce additional uncertainty (and potentially a net bias) to 
any lensing-based estimates of the cluster mass contained within a 3D volume. 
Studies based on simulated clusters (e.g., Clowe et al.\ 2004) 
indicate that the net bias is no more 
than a few percent when 3D cluster masses are estimated by fitting observations of  
$g_T (r)$ to predictions from theoretical models of the mass 
distribution, such as the NFW model. 
However, the scatter in the mass estimates from projection effects 
amount to a weak lensing mass dispersion of $\sim 15-25\%$ for massive 
galaxy clusters, which should be added to the observational uncertainties of the 
lensing mass estimates.   
In this paper, we have assumed a lensing mass dispersion of 0.26 resulting from projection effects, 
corresponding to the value estimated by Metzler, White \& Loken (2001) from their N-body simulations.
Although these authors considered a somewhat different mass estimator, more recent estimates 
indicate a similar mass dispersion for the NFW profile fitting method that we have used.    
This additional mass uncertainty has been added in quadrature to the uncertainties of $M_{500c}$
values listed in Table~\ref{tab:dataset}. 

The absence of reliable information about the individual redshifts of the faint galaxies used 
for the weak lensing measurement will inevitably result in some degree of confusion between 
lensed background galaxies and unlensed cluster galaxies. The magnitude of this effect will 
depend on the projected number density of cluster galaxies, 
and should thus have a strong dependency on cluster radius. Hence, a radially dependent correction 
factor was applied to the shear measurements to correct for  
contamination from cluster galaxies in the faint galaxy catalogs that were used to 
measure the gravitational shear (these catalogs included all galaxies in the cluster fields
that were detected at a signal-to-noise ratio $6 < S/N < 100$, with no additional selection
based on e.g., galaxy color).  
The magnitude of this correction was estimated from the radial dependence of the average faint 
galaxy density in two ``stacks'' of clusters observed with the UH8K camera, 
one at $z\sim 0.30$ and the other at $z\sim 0.23$, assuming that the 
contamination is negligible at the edge of the UH8K fields, $>1.5h^{-1}$ Mpc from the cluster center.
The estimated degree of contamination is shown in Figure~\ref{fig:cluscont}. 

Given the difference in cluster redshift, the similarity of the two curves in 
Figure~\ref{fig:cluscont} may be somewhat surprising, as one would naively expect the 
more distant clusters to display a significantly lower surface density of cluster 
galaxies. 
However, there are several competing effects that affect the observed galaxy 
density at a fixed angular radius: Firstly, if all cluster galaxies were detectable 
regardless of cluster redshift, the change in apparent image scale with redshift should 
increase the surface density by a factor given by the square of the ratio of the angular 
diameter distances. On the other hand, a fixed angular radius would 
correspond to a larger physical cluster-centric radius (and hence lower galaxy density in 
physical units) at the larger redshift, the difference depending on the slope of the 
radial galaxy density profile. At the radii probed in this study, both the radial 
surface mass density profile and the number density profile of bright cluster galaxies 
follow approximately the power law behavior of a singular isothermal sphere 
($\sigma (r) \propto r^a$, with $a=-1$). Hence, the physical number density 
(in galaxies/Mpc$^2$) at a fixed angular radius should decrease as the inverse of the 
ratio of angular diameter distances. In addition, the faintest galaxies drop below
the detection limit at higher redshift, the effect depending on the slope of the cluster 
luminosity function around $M_R \approx -15$. Assuming a Schechter (1976) luminosity function 
with a faint-end slope $\alpha^{\star} = -1.25$ (typical of rich clusters) and 
$M_R^{\star} (z=0.23) = -21.65$, 
the luminosity function can be integrated down to the detection limit (corresponding 
to $M_R \simeq -15.0$ and $M_R \simeq -15.7$ at $z=0.23$ and $z=0.30$, respectively),
to estimate the fraction of cluster galaxies that drop out at the higher 
redshift ($\sim 30$\%).   
Finally, a redshift-dependence given by $M_R^{\star} (z) = M_R^{\star} (0) + 5\log (1+z)$ 
was assumed to account for galaxy evolution in the clusters. The combination of all these 
effects would predict a surface density of cluster galaxies which is 7\% less at 
$z=0.3$, compared to $z=0.23$, for a fixed cluster richness. 
Even this small difference would be erased by a slight decrease in  
the assumed values of the slopes $\alpha^{\star}$ and $a$.  
A faint-end slope of the luminosity 
function of $\alpha^{\star} = -1.1$ would be sufficient to remove 
the predicted difference in galaxy surface 
density at the two different redshifts. Based on {\it Hubble Space Telescope} (HST) WFPC2 
imaging of the galaxy cluster \objectname{A2218} (which is similar to the clusters studied here in 
terms of optical richness, lensing mass and X-ray properties), Pracy et al.\ (2004) 
find that the cluster core shows a relative depletion of dwarf 
galaxies, leading to a radial profile of faint galaxies which is significantly shallower 
than the SIS prediction. For the ``intermediate'' dwarf population 
($-18 < M_{\rm F606W} < -15$, similar to the range in absolute magnitude of cluster 
galaxies in our faint galaxy catalogs), these authors find a radial distribution with a 
slope $a=-0.63\pm0.09$. Assuming a similar slope for the radial distribution of the 
faint cluster galaxies in our catalogs 
would also remove the predicted difference between the two curves in 
Figure~\ref{fig:cluscont}. Based on this figure, 
and the above discussion, we conclude that we are probably justified in ignoring the 
redshift-dependence in our cluster galaxy contamination correction.      
   
The level of cluster galaxy contamination for individual clusters will generally differ 
from the mean level calculated above, as there will be significant cluster-to-cluster variations 
in the abundances of cluster dwarf galaxies. Based on the sample of clusters observed with the 
UH8K camera, 
the scatter in dwarf galaxy richness was estimated to be $\sim 50\%$ (this estimate also includes 
variations in the field galaxy density caused by uncorrelated large-scale structures along the line of 
sight, and hence the true scatter in dwarf galaxy richness of the clusters is somewhat 
overestimated). By employing the mean contamination correction calculated above rather than an 
estimate appropriate for each cluster, we introduce an additional scatter of up to $20\%$ in our     
mass estimates. This additional scatter is not included in the uncertainties of the 
tabulated mass measurements in Table~\ref{tab:dataset},  but is considered further in
Section~3.2.

\begin{figure}
\resizebox{8cm}{!}{\includegraphics{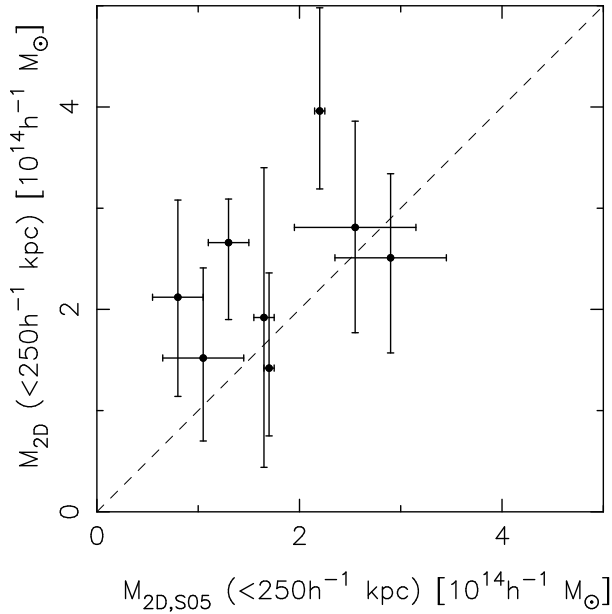}}
\caption[Comparison with S05 mass estimates.]
{Comparison with estimates of the projected mass within $250 h^{-1}$ kpc published by 
S05. The lensing masses in this plot are given for the 
cosmological model assumed by S05, and our estimate of 
the degree of galaxy cluster contamination was also adjusted to match the 
contamination assumed by S05 (see text for details). }
\label{fig:S05comp}
\end{figure}

Eight of the clusters in our sample were also included in the combined strong and weak gravitational 
lensing study of S05, based on observations of a sample of 10 X-ray luminous galaxy clusters at 
$z\sim 0.2$ using HST WFPC2. 
These authors estimated the projected cluster mass within a clustercentric radius of 
$250 h^{-1}$ kpc, assuming an Einstein-de Sitter ($\Omega_m = 1$, $\Omega_{\Lambda} = 0$) cosmology. 
Figure~\ref{fig:S05comp} shows a comparison of the mass values listed by S05 with our 
 cluster mass estimates, using the best-fit NFW model to derive projected cluster masses, assuming 
the same cosmology as S05. These authors assumed a spatially constant contamination of 
20\% cluster galaxies in their background galaxy catalogs at radii $<2\arcmin$, while we find an 
average contamination of 30\% for our data at these radii. 
Hence, for the plot in Figure~\ref{fig:S05comp} we 
have adjusted our radially dependent contamination correction such that the average contamination 
at small radii is consistent with that assumed by S05. We find that our cluster mass estimates are 
generally consistent with those of S05, although with a tendency for higher masses 
(by about 30\%). 
     
\subsection{X-ray data}

For clusters in the weak lensing data set, we compiled a list of 
corresponding X-ray temperatures from the literature. 
For many of these clusters, their global temperature,
or even a temperature map, has been determined using data 
from {\it Chandra} and/or XMM-Newton.
However, these temperatures constitute a rather heterogeneous sample,
for which the systematics are not well established.
Consequently, X-ray temperatures
were primarily drawn from the samples of \cite{OM}, \cite{allen00} and \cite{white00}, 
each providing a homogeneous measure of the global cluster temperature 
(i.e., temperature measured within a cluster-centric distance close to 
$r_{500c}$) for a large fraction of the clusters in the weak lensing sample. 
All these authors derived temperatures based on analysis of ASCA spectra. 
For two clusters not in either of the samples mentioned above
we extracted published temperatures from other sources.

Specifically, from the works of \cite{OM} and \cite{allen00} we
extracted temperatures estimated by these authors by fitting an isothermal plasma model 
with the Galactic absorbing column density as a free parameter. 
The temperatures taken from \cite{white00} were derived by fitting an isothermal plasma
model with the nominal Galactic absorbing column density fixed, but since only
energies above 1~keV are used in the \cite{white00} spectral fitting the
fixed column density should not
introduce systematic effects relative to the temperatures from the 
\cite{OM} and the \cite{allen00} samples.
For the remaining two clusters we 
extracted published temperatures obtained in a similar way 
(see Table~\ref{tab:dataset}).

For the clusters in two or more samples, their derived temperatures agree
within the uncertainties, and for any cluster the derived temperatures 
differ by less than 20\% between samples. Also, the mean of temperature 
differences between
any two of the samples by \cite{OM}, \cite{allen00}, and \cite{white00} is 
less than 3\%, indicating the low level of systematic temperature variance between 
different analyses.  

Although the isothermal plasma model has proven too simplistic for nearby 
clusters, the global cluster temperature is straightforward
to derive from observations as well as simulations, enabling a rather
direct comparison between observations and theory. 
Furthermore, for the majority of distant ($z \gtrsim 0.5$) clusters only global 
isothermal temperatures can be obtained in the foreseeable future. 
Hence, we refrain from going into the detailed spatial and
spectral modeling of the intra-cluster gas. 
The effects of cluster dynamics, ``cooling 
cores'', non-sphericity etc. generally affects the global temperatures
only at the 10\%-20\% level \citep[e.g.,][]{evrard96,white00,smith}.

\begin{deluxetable*}{lcccccccc}
\tablecaption{Weak lensing masses and X-ray temperatures 
	\label{tab:dataset}}
\tablehead{
\colhead{Cluster\hspace*{20mm}} &
\colhead{$z_{\rm cl}$\rlap{\tablenotemark{a}}} &
\colhead{$M_{500c}$} &
\colhead{$r_{\rm fit}/r_{500c}$} &
\colhead{$kT$ (Ota \& Mitsuda)} &
\colhead{$kT$ (Allen)} &
\colhead{$kT$ (White)} &
\colhead{$kT$ (other)\rlap{\tablenotemark{b}}} &
\\
 & & \colhead{($10^{14}\, h^{-1} M_{\sun}$)} & & \colhead{(keV)} & \colhead{(keV)} & \colhead{(keV)} & \colhead{(keV)}
}
\startdata
\objectname{A68} \dotfill  & 0.255 &   $21.37^{+6.79}_{-7.20}$ & 0.30 & $6.93^{+0.63}_{-0.59}$ & \nodata & \nodata & \nodata \\
\objectname{A115} \dotfill  & 0.197 &   $2.42^{+2.34}_{-1.84}$  & 0.50 &  $5.83^{+0.47}_{-0.30}$ & \nodata & \nodata & \nodata \\
\objectname{A209} \dotfill  & 0.206 &   $7.54^{+4.24}_{-3.87}$  & 1.62 & \nodata                & \nodata & \nodata & $7.10^{+0.40}_{-0.40}$ \\
\objectname{A267} \dotfill  & 0.230 &   $8.79^{+2.87}_{-3.53}$  & 1.32 & $5.51^{+0.44}_{-0.41}$ & \nodata & \nodata & \nodata \\
\objectname{A520} \dotfill  & 0.203 &   $8.67^{+3.36}_{-2.60}$  & 1.20 & \nodata                & $7.94^{+0.96}_{-0.90}$ & \nodata & \nodata \\
\objectname{A586} \dotfill  & 0.171 &   $25.27^{+7.01}_{-8.11}$ & 0.21 & $6.96^{+0.99}_{-0.83}$ & $7.02^{+0.94}_{-0.80}$ & $6.06^{+0.64}_{-0.52}$ & \nodata \\
\objectname{A611} \dotfill  & 0.288 &   $3.83^{+2.99}_{-2.79}$  & 0.59 & \nodata                & \nodata & $6.85^{+0.48}_{-0.46}$ & \nodata \\
\objectname{A665} \dotfill  & 0.182 &   $5.40^{+3.40}_{-3.07}$  & 0.36 & $6.96^{+0.28}_{-0.27}$ & $8.12^{+0.62}_{-0.54}$ & $7.73^{+0.41}_{-0.35}$ & \nodata \\
\objectname{A697} \dotfill  & 0.282 &   $12.18^{+4.97}_{-4.89}$ & 0.39 & $8.19^{+0.62}_{-0.60}$ & \nodata & $8.60^{+0.50}_{-0.49}$ & \nodata \\
\objectname{A773} \dotfill  & 0.217 &   $13.09^{+4.79}_{-6.15}$ & 0.30 & $8.07^{+0.70}_{-0.66}$ & $8.29^{+0.73}_{-0.64}$ & $8.63^{+0.68}_{-0.67}$ & \nodata \\
\objectname{A959} \dotfill  & 0.285 &   $9.33^{+3.50}_{-3.05}$  & 1.45 & $5.24^{+0.89}_{-0.73}$ & \nodata & \nodata & \nodata \\
\objectname{A963} \dotfill  & 0.206 & $4.42^{+4.27}_{-3.46}$  & 1.51 & $6.83^{+0.51}_{-0.51}$ & $6.13^{+0.45}_{-0.30}$ & $6.08^{+0.43}_{-0.33}$ & \nodata \\
\objectname{A1576} \dotfill & 0.299 & $8.62^{+3.40}_{-2.54}$  & 1.58 & \nodata                & \nodata & $6.57^{+0.56}_{-0.54}$ & \nodata \\
\objectname{A1682} \dotfill & 0.226 & $2.24^{+1.9}_{-1.33}$   & 0.55 & $6.42^{+0.63}_{-0.60}$ & \nodata & $7.24^{+0.68}_{-0.59}$ & \nodata \\
\objectname{A1722} \dotfill & 0.325 & $2.70^{+2.14}_{-1.58}$  & 2.58 & $5.81^{+0.59}_{-0.39}$ & \nodata & \nodata & \nodata \\
\objectname{A1758N} \dotfill & 0.280 & $20.37^{+6.37}_{-6.65}$ & 0.33 & $6.88^{+0.86}_{-0.75}$ & \nodata & \nodata & \nodata \\
\objectname{A1763} \dotfill & 0.228 & $4.90^{+2.42}_{-3.07}$  & 0.45 & $8.11^{+0.66}_{-0.63}$ & \nodata & $7.30^{+0.46}_{-0.38}$ & \nodata \\
\objectname{A1835} \dotfill & 0.253 & $8.42^{+4.41}_{-3.33}$  & 0.39 & $7.42^{+0.61}_{-0.43}$ & $7.33^{+0.35}_{-0.30}$ & $7.88^{+0.49}_{-0.46}$ & \nodata \\
\objectname{A1914} \dotfill & 0.171 & $2.62^{+2.03}_{-1.93}$  & 0.44 & \nodata                & \nodata & $10.53^{+0.51}_{-0.50}$ & \nodata \\
\objectname{A1995} \dotfill & 0.320 & $23.69^{+6.88}_{-6.13}$ & 1.23 & $9.06^{+1.77}_{-1.32}$ & \nodata & $7.57^{+1.07}_{-0.76}$ & \nodata \\
\objectname{A2104} \dotfill & 0.153 & $14.14^{+5.34}_{-5.55}$ & 0.23 & $7.66^{+0.49}_{-0.43}$ & \nodata & $9.12^{+0.48}_{-0.46}$ & \nodata \\
\objectname{A2111} \dotfill & 0.229 & $3.84^{+1.80}_{-2.14}$  & 0.47 & $6.94^{+0.76}_{-0.67}$ & \nodata & \nodata & \nodata \\
\objectname{A2204} \dotfill & 0.152 & $7.86^{+5.28}_{-4.57}$  & 0.27 & $6.68^{+0.28}_{-0.27}$ & $6.23^{+0.30}_{-0.28}$ & $6.99^{+0.24}_{-0.23}$ & \nodata \\
\objectname{A2219} \dotfill & 0.228 & $4.72^{+2.35}_{-2.94}$  & 0.45 & \nodata                & $9.46^{+0.63}_{-0.57}$ & $9.52^{+0.55}_{-0.40}$ & \nodata \\
\objectname{A2261} \dotfill & 0.224 & $11.52^{+5.40}_{-5.97}$ & 0.32 & $6.56^{+0.49}_{-0.48}$ & $6.64^{+0.51}_{-0.46}$ & $7.49^{+0.57}_{-0.43}$ & \nodata \\
\objectname{MS1455+22} \dotfill & 0.258 & $3.21^{+2.19}_{-1.78}$ & 0.58 & \nodata                & $4.33^{+0.27}_{-0.25}$ & $4.83^{+0.22}_{-0.21}$ & \nodata \\
\objectname{RX\,J1532.9+3021} \dotfill & 0.345 & $13.86^{+5.93}_{-5.64}$ & 0.44 & $4.91^{+0.29}_{-0.30}$ & \nodata & \nodata & \nodata \\
\objectname{RX\,J1720.1+2638} \dotfill & 0.164 & $3.28^{+2.65}_{-2.53}$ & 0.39 & \nodata                & \nodata & \nodata & $5.60^{+0.50}_{-0.50}$ \\
\objectname{RX\,J2129.6+0005} \dotfill & 0.235 & $8.35^{+4.62}_{-4.87}$ & 0.38 & $5.72^{+0.38}_{-0.30}$ & \nodata & \nodata & \nodata \\
\objectname{Zw3146} \dotfill & 0.291 & $7.57^{+4.13}_{-3.25}$ & 0.47 & \nodata                & $6.80^{+0.38}_{-0.36}$ & $5.89^{+0.30}_{-0.22}$ & \nodata \\
\enddata
\tablenotetext{a}{--- See Ebeling et al.\ (1996; 1998)
 for references to redshift measurements.} 
\tablenotetext{b}{--- $kT$ value for \objectname{A209} from P.B.\ Marty 
 (private communication); $kT$ value for \objectname{RXJ1720.1+2638} from \citet{mazzotta}.
}
\end{deluxetable*}

\section{Results}
\label{sec:results}
From the virial relation
\begin{equation} 
M \propto \langle v^2 \rangle r \propto T r,
\end{equation} 
between cluster mass, $M$, inside radius $r$, galaxy velocity, $v$, and gas temperature, $T$, 
combined with the definition of mass within an over-density of 500 times the critical density 
\begin{equation} 
M_{500c} \propto r_{500c}^3 \rho_c(z) \propto r_{500c}^3 E^2(z) \rho_c(0) \propto r_{500c}^3 E^2(z),
\end{equation}
where the term $E^2(z)=\rho_c(z)/\rho_c(0)$ describes the evolution of the over-density for a given
cosmology, the mass--temperature relation is obtained as
\begin{equation} 
E(z)M_{500c} \propto T^{3/2}.
\label{eq:theoryMT}
\end{equation} 

In this study, we have measured $M_{500c}$, relating to $M_{\rm vir}$ through 
$M_{\rm vir} = 1.65 M_{500c}$ for the average cluster redshift with
our chosen NFW model. We take into account that the slope may deviate from the simple theoretical 
expectation $\alpha = 3/2$ and normalize the relation at 8~keV since our sample 
is dominated by massive clusters. Hence, mass--temperature relations were obtained by fitting 
the data in Table~\ref{tab:dataset} using the BCES($X_2|X_1$) estimator 
of \cite{ab} to the following parameterization of the mass--temperature relation

\begin{equation} 
E(z)M_{500c}= M_{\rm 500c, 8 keV} (T/8 {\rm keV})^{\alpha}.
\label{eq:fittedMTrelation}
\end{equation} 

The redshift-dependent factor, $E(z)$, contained in
eq.\ref{eq:fittedMTrelation} must be calculated individually for 
each cluster, as this would otherwise 
produce an artificial 15\% variation in mass over the redshift 
range spanned by our cluster sample. In effect, the normalization of the relation refers
to the present epoch ($z=0$).

\begin{figure}
\resizebox{9cm}{!}{\includegraphics{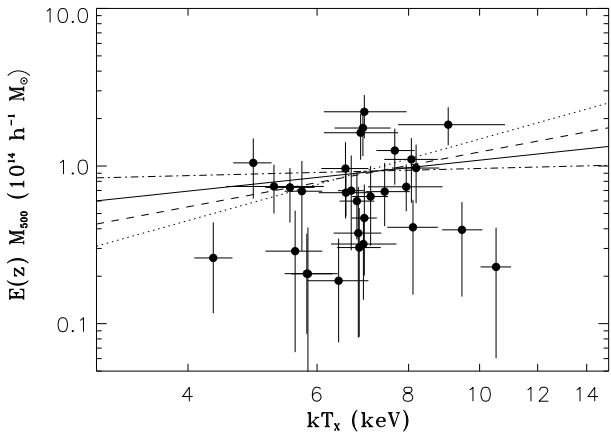}}
\caption[Mass--temperature relation.]
{Weak lensing mass estimates and X-ray temperatures with BCES($X_2|X_1$)
regression lines for different subsamples. The full line is
the fit to the full sample, defined as described in the text; the dotted line
is the fit to the \cite{OM} sample; the dashed line is the fit to the \cite{allen00}
sample; and the dot--dashed line is the fit to the \cite{white00} sample.}
\label{fig:MTrel}
\end{figure}

The fitting procedure of \cite{ab} takes uncertainties in temperatures as well 
as in weak lensing masses into account, and makes no assumptions about
the intrinsic scatter of both quantities. Results from fitting sub-samples
as well as the full sample are presented in Table~\ref{tab:Tstar} and 
Figure~\ref{fig:MTrel}.

For the full data set (for those clusters with temperature from more than one
sample the temperature was taken in prioritized order from \cite{OM}, \cite{allen00},
and \cite{white00}) we find the following normalization of the mass--temperature 
relation at 8 keV 
$M_{\rm 500c,8 keV}=(8.7 \pm 1.6)\times {10^{14} h^{-1} M_{\sun}}$ and a slope of $\alpha = 0.49 \pm 0.80$.
It is evident that the slope of the mass--temperature 
relation is not well-determined since our data only span a modest 
range at the high mass/high temperature end of the cluster distribution.
In fact, it is not obvious that there is a tight mass--temperature relation
at the high temperature end probed here.

\subsection{Normalization of the mass--temperature relation and $\sigma_8$}
The concentration of clusters around $6-8$~keV enables a robust 
measurement of the normalization of the mass--temperature relation
at the high--mass end. Even though the slope of the mass--temperature
relation varies substantially between the three X-ray sub-samples 
\citep{OM, allen00, white00} the best fit normalizations agree within their statistical uncertainty. 
We note that for all fits, the four different regressions
of \cite{ab} all result in normalizations within 20\%.

The strongest constraints on the mass--temperature 
relation normalization are obtained by taking advantage of previous
studies of massive clusters \citep{fin,allen01,arnaud05}, showing that the
mass--temperature relation slope is close to $\alpha=3/2$ as expected from simple 
gravitational collapse models \citep{kaiser}.
Hence, in order to express the normalization of the mass--temperature relation in terms
of the characteristic temperature $T_{\star}= 8{\rm keV}\left(1.65 M_{\rm 500c,8 keV}\right)^{-1/\alpha}$ 
\citep{pier03} we assume $\alpha = 3/2$ \citep[with a representative uncertainty of 10\%, e.g.][]{fin}. 
As is custom for quoting $T_{\star}$ values we adopt the redshift dependence factor 
$F(z)=\left(\Delta_c E^2\right)^{-1/2} \left[ 1- 2\frac{\Omega_{\Lambda} (z)}{\Delta_c} \right]^{-3/2}$
\citep{pier03} where $\Delta_c$ is the mean overdensity inside the virial radius in units 
of the critical density at the relevant redshift ($F(z)$ and $E(z)$ differs by 7\% at $z=0.23$).

We find $T_{\star} = 1.28\pm 0.20$ for our full sample and results from calculations of 
$T_{\star}$ based on various subsamples are listed in Table~\ref{tab:Tstar2}. 
From the $\sigma_8$ - $T_{\star}$ relation plotted by \cite{pier03} 
in their Figure 2, we find $\sigma_8 = 0.88\pm 0.09$, based on our full sample. 
We note that this relation is 
valid only for an intrinsic scatter in temperature of $\lesssim 10\%$ 
around the mean mass--temperature relation. A larger intrinsic scatter will imply a lower 
value of $\sigma_8$. We provide our constraints on the intrinsic scatter below.

\subsection{Scatter in the mass--temperature relation}
The squared scatter in lensing mass, $M_{500c}$, around the best fit, 
$(\sigma_M^{tot})^2=\langle (\frac{M_{500c}^i-M_{500c}^{fit}}{M_{500c}^{i}})^2 \rangle$,
is the sum of the squared measurement error, the squared intrinsic scatter, and the
squared systematic errors
$(\sigma_M^{tot})^2=(\sigma_M^{err})^2+(\sigma_M^{i})^2+(\sigma_M^{sys})^2$. 
The main systematic errors in the lensing mass (see Section~2.1) arise from extrapolating the 
assumed NFW mass profile out to $r_{500c}$
(due to cluster-to-cluster variations in the assumed concentration parameter 
$c_{vir}$) and from the separation of cluster/background galaxies (due to 
cluster-to-cluster richness variations). Each of these introduces a scatter 
of 20\%{} in the lensing mass, hence $\sigma_M^{sys}=0.28$. For the full sample
we find $\sigma_M^{tot}=0.94$ which is larger than expected from the mean lensing 
mass error, $\sigma_M^{err}=0.52$ and the systematic errors, indicating either a sizable 
intrinsic scatter in mass or that the measurement/systematic errors are severely under-estimated. 

Accounting for errors in both mass and temperature, we find an 
intrinsic scatter in $T_{\star}$ of $0.25^{+0.28}_{-0.25}$. 
There is a 70\% probability that the scatter in temperature is larger than 10\%, 
favoring somewhat lower values of $\sigma_8$ than quoted above.  
However, most of the scatter is caused by the low mass clusters. 

\subsection{Relaxed vs. non-relaxed clusters}
\begin{figure}
\resizebox{9cm}{!}{\includegraphics{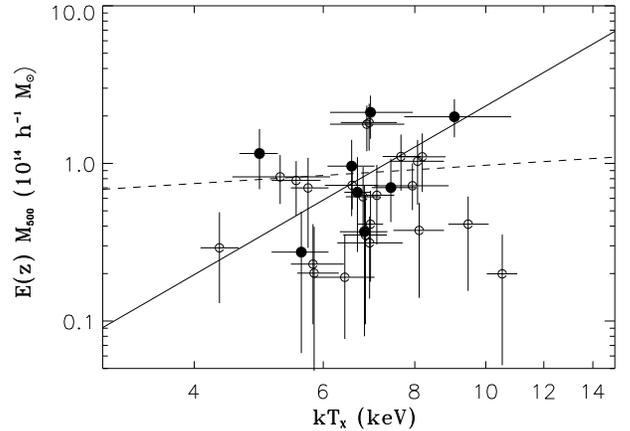}}
\caption[Mass--temperature relation.]
{Weak lensing mass estimates and X-ray temperatures with BCES($X_2|X_1$)
regressions for ``relaxed'' clusters (filled symbols, full line) 
and ``non-relaxed'' clusters (open symbols, dashed line).}
\label{fig:relax}
\end{figure}

We looked into whether relaxed clusters and non-relaxed clusters have the
same normalization of the mass--temperature relation. Our ``relaxed'' 
cluster sample consists of A586, A963, A1835, A1995, A2204, A2261, RXJ1720, 
and RXJ1532. These are clusters with ``spherical'' optical and X-ray 
morphology, and no known cluster-scale dynamic disturbances. 
For the relaxed clusters we find a normalization
of the mass--temperature relation of 
$M_{\rm 500c, 8 keV, relax}=(17.3\pm 3.7)\times {10^{14} h^{-1} M_{\sun}}$ 
while the normalization for the non-relaxed clusters is 
$M_{\rm 500c, 8 keV, nonrelax}=(7.6\pm 1.5)\times {10^{14} h^{-1} M_{\sun}}$
(see Figure~\ref{fig:relax}). 
The higher normalization of relaxed clusters
is supported by the fact that the mean mass of relaxed clusters is a factor
$1.5$ larger than the mean mass of ``non-relaxed'' clusters, although
the relaxed and the non-relaxed clusters
span roughly the same temperature range.

The scatter in mass for the relaxed sample ($\sigma_M^{tot}=0.77$) is
similar to the scatter for the non-relaxed sample ($\sigma_M^{tot}=0.88$). 
The mean error for both samples is $\sigma_M^{err}=0.57$. Either relaxed 
clusters spread as much around their mass--temperature relation as 
clusters in general, or we have used a poor definition of ``relaxed'' clusters.
However, the fact that the normalization of the mass--temperature relation for
relaxed clusters is higher than for non-relaxed clusters
indicates that there is a physical difference between the two sub-samples.
From the present study, it thus seems that relaxed clusters do not form a 
tighter mass--temperature relation than clusters in general.

For a given mass, non-relaxed clusters are found to be $\sim 75\pm 40$\% hotter than
relaxed clusters. Since we consider global, isothermal temperatures, 
the presence of ``cooling cores''
in relaxed clusters will result in a lower global temperature than
the virial temperature. However, this effect is at the 10\%-20\% level
\citep[e.g.,][]{smith03, ohara} so this cannot alone explain the temperature 
difference between
relaxed and non-relaxed clusters. Based on the mass--temperature relation from 
10 clusters (3 of which are considered relaxed), S05 
also find that non-relaxed clusters are hotter 
than relaxed clusters. An objective classification of the degree of relaxation
for a sizeable cluster sample is required for further quantifying
the size of this effect. 

\begin{deluxetable}{ccr}
\tablecaption{Best-fit mass-temperature relation (arbitrary slope)
	\label{tab:Tstar}}
\tablehead{
\colhead{$M_{\rm 500c, 8 keV}$} & 
\colhead{$\alpha$} & 
\colhead{Sample}\\
\colhead{$(10^{14} h^{-1} M_{\sun})$} &
\colhead{} &
\colhead{}\\
}
\startdata
$12.0\pm 2.4$   & $1.30\pm 0.97$ & Ota \& Mitsuda (2004) \\
$9.5\pm 1.9$   & $0.87\pm 0.78$ & Allen (2000) \\
$8.1\pm 1.5$   & $0.11\pm 0.98$ & White (2000) \\
$8.7\pm 1.6$   & $0.49\pm 0.80$ & All \\
$17.3\pm 3.7$   & $2.69\pm 1.30$ & ``Relaxed'' \\
$7.6\pm 1.5$   & $0.29\pm 0.76$ & ``Non-relaxed'' \\
\enddata
\end{deluxetable}

\begin{deluxetable}{ccr}
\tablecaption{Best-fit mass-temperature relation (fixed slope)
	\label{tab:Tstar2}}
\tablehead{
\colhead{$T_{\star}$} & 
\colhead{$\alpha$} & 
\colhead{Sample}
}
\startdata
$1.04\pm 0.18$ & $3/2$ & Ota \& Mitsuda (2004) \\
$1.22\pm 0.21$ & $3/2$ & Allen (2000) \\
$1.35\pm 0.21$ & $3/2$ & White (2000) \\
$1.28\pm 0.20$ & $3/2$ & All \\
$0.82\pm 0.14$ & $3/2$ & ``Relaxed'' \\
$1.42\pm 0.22$ & $3/2$ & ``Non-relaxed'' \\
\enddata
\tablecomments{The uncertainty in $T_{\star}$ from 
$\alpha$ has been taken to be 10\%.}
\end{deluxetable}

\begin{deluxetable*}{lcccr}
\tablecaption{Normalizations of the mass-temperature relation
           \label{tab:Mnorm}}
\tablehead{
\colhead{Method} & 
\colhead{$\langle z \rangle$} & 
\colhead{$M_{\rm 500c,8 keV}$} & 
\colhead{Slope} & 
\colhead{Reference}
}
\startdata
X-rays        & 0.09 & $6.00\pm 0.35$   & Fitted & \citet{arnaud05} \\
                  & 0.09 & $6.07\pm 0.46$   & Fitted & \citet{vik} \\
     		    &  &  &  \\ 
Lensing       & 0.23 & $6.65\pm 0.52$   & Fixed & S05 \\
                  & 0.23 & $8.67\pm 1.57$   & Fitted & This study \\
                  &  &  &  \\ 
Simulations & 0.04 & $8.09\pm 0.48$   & Fixed  & \citet{evrard96} \\
                  & 0.00 & $6.82\pm 0.76$   &  Fitted & \citet{borgani} \\
\enddata
\tablecomments{Included here are some recent studies with a quoted 
normalization of the $M_{500c}-T$--relation, scaled to $M_{\rm 500c,8 keV}$, 
i.e., the mass contained within $r_{500c}$ of a cluster with $kT = {\rm 8 keV}$, 
with errors propagated, and assuming $h=1$. The listed normalization has been scaled to a 
common redshift of $z=0$.  
The normalization of S05, which was 
determined within a fixed physical radius of $250 h^{-1}$~kpc in an Einstein-de Sitter universe,   
has been scaled to $M_{500c}$ for our adopted cosmology, making the same 
assumptions about an NFW-type mass profile as we have made for our own data. 
For each study, we indicate whether the slope of the relation 
was calculated from a fit to the data, or whether the slope was 
fixed at the theoretically expected value, $\alpha=3/2$.
}
\end{deluxetable*}

\section{Discussion}
\label{sec:discussion} 
Based on the hitherto largest sample of X-ray luminous clusters with measured
lensing masses, we derive a normalization of the mass--temperature 
relation at the high mass end, $M_{\rm 500c, 8 keV}=(8.7\pm 1.6)\,h^{-1}\,10^{14}M_{\sun}$. 
This value is higher than the lensing based mass--temperature normalization of S05, 
based on a smaller cluster sample, but is consistent with this 
within $1\sigma$ errors; see Table~\ref{tab:Mnorm}.
Mass--temperature relations with masses determined from X-ray data tend to have
a lower normalization than lensing based relations, and they are only 
marginally consistent with our normalization. This is also the case for
the two recent studies of \citet{vik} and \citet{arnaud05} based on smaller
samples of lower mass (and hence cooler) clusters.
\citet{vik} measured cluster masses inside $r_{500c}$ from X-ray observations 
of a sample of 13 low redshift clusters with a median temperature of 5.0~keV while
\citet{arnaud05} determined the normalization
from X-ray derived masses of 10 nearby clusters with a mean temperature of 4.8~keV.
The two studies agree on the same normalization, higher than previous
X-ray mass based studies, but there still seems to be a $\sim 20\%$ discrepancy
between X-ray and lensing derived mass--temperature relations.

We note, however, that the lensing based and X-ray based normalizations are made at 
different redshifts, and that this discrepancy would vanish if the redshift-dependence 
predicted by the self-similar collapse model in equation~\ref{eq:theoryMT} were neglected.  
Given the heterogeneous nature of these data sets, any claim of significant departures from
self-similarity would be premature, but this clearly provides an interesting avenue for 
future research, involving even larger cluster samples spanning a wider interval in redshift.
  
We confirm the result of Smith et al.\ (2005) that non-relaxed clusters are 
on average significantly hotter than relaxed clusters. This is qualitatively consistent with
N-body/hydrodynamical cluster simulations which show that major mergers can temporarily 
boost the X-ray luminosities and temperatures well above their equilibrium values
\citep[e.g.][]{randall}.   

In contrast to several previous (mainly X-ray mass based) published mass--temperature
relations, the normalization derived in this study is in good agreement with 
the normalization derived from numerical simulations. However,
the accuracy of the normalization is not good enough to discriminate between
simulations including different physical processes. 
Our results show that X-ray based measurements of the cluster abundances, after  
reducing the major systematic uncertainties associated with the mass--temperature 
normalization, give an amplitude of mass fluctuations on cluster scales that 
is consistent with other methods.
This lends additional support to the ``concordance model'' cosmology, and lends 
credence to the basic assumptions of Gaussian density fluctuations. 
Our determination of $\sigma_8=0.88\pm 0.09$ is higher than most
$\sigma_8$ determinations from cluster data \citep[for a compilation of these, 
see e.g.,][]{henry04}.
However, our finding is consistent with the value derived from weak 
gravitational lensing in the combined Deep and Wide CFHT Legacy Survey
($\sigma_8 = 0.86\pm 0.05$; Semboloni et al.\ 2005) based on the 
halo model of density fluctuations (Smith et al.\ 2003b). It is also  
consistent with the CMB+2dFGRS+Ly$_{\alpha}$ forest 
result ($\sigma_8 = 0.84\pm 0.04$) of Spergel et al.\ (2003), with 
the joint CMB + weak lensing analysis of Contaldi, Hoekstra, \& Lewis (2003), 
which gave $\sigma_8 = 0.89\pm 0.05$, and with CMB analyses \citep{bond}
yielding $\sigma_8 \approx 0.9$. However, the more recent 3-year WMAP  
results (Spergel et al.\ 2006) give a significantly lower value of $\sigma_8$, 
and also a preference for a value of $\Omega_m$ lower than 0.3. 
Also, results from the recent 100 square degree weak lensing survey (Benjamin et al.\ 2007) 
favor a lower value of $\sigma_8 = 0.74 \pm 0.06$ for $\Omega_m = 0.3$.
We note that our quoted value of $\sigma_8$ is based on the assumption
that the intrinsic scatter about the mass--temperature relation is $\lesssim 10\%$, 
and that our $\sigma_8$ estimate will be biased high if the true scatter significantly 
exceeds this value \citep{pier03}.  

The limiting factor of our measurement of the normalization of the 
mass--temperature relation is the magnitude of the measurement errors
(dominating the systematic errors, estimated to be $\sim 30\%$).
In order for the mass--temperature relation to be a competitive route
for constraining cosmological parameters and to discriminate
between simulations with different input physics, the normalization must be
measured to better than $\sim 10$\% accuracy. 
However, there are good prospects for improving on these results in the near future. 
Firstly, the superior spectro-imaging capabilities of {\it Chandra} and XMM-Newton 
will allow the construction of large, homogeneous cluster temperature samples.
A comparison to tailored simulations
with realistic physics, analyzed in the same way as observations, will
advance our understanding of systematics and the link between the mass--temperature 
relation and structure formation (C.B.\ Hededal et al.\ 2007, in preparation). 
Secondly, more accurate weak lensing-based 
mass measurements of a larger sample of clusters are feasible as large mosaic CCD   
cameras that can probe intermediate-redshift clusters beyond their virial radii are 
now common, and the cluster sample could easily be doubled from a similar survey 
in the Southern celestial hemisphere.  

Finally, we note that a more direct measurement of $\sigma_8$ from weak lensing by clusters is 
possible, provided that weak lensing mass estimates are available for a large, 
well-defined, volume-limited cluster sample. This makes it feasible to calculate the cluster mass 
function directly from the lensing masses, rather than indirectly via the X-ray temperature function   
(Dahle 2006). Since mass estimates based on baryonic tracers of the total cluster mass 
 only enters indirectly as a selection criterion (e.g., clusters selected based on X-ray luminosity 
above a certain threshold), the method is less susceptible to systematic and 
random errors, as it does not require an accurate characterization of the 
scatter around the mean mass--temperature relation.  

\acknowledgments

We thank Per B. Lilje and Jens Hjorth for valuable comments on a draft version of 
this paper. We also thank the anonymous referee for comments and suggestions
that improved the presentation of our results.
KP acknowledges support from the Danish National Research Council, the
Carlsberg foundation, and Instrument Center for Danish Astrophysics.
The Dark Cosmology Centre is funded by the Danish National Research Foundation.
HD acknowledges support from the Research Council of Norway and travel support 
from NORDITA.

\end{document}